\documentclass[%
reprint,
showpacs,
 amsmath,amssymb,
 aps,
pra,
]{revtex4-1}

\usepackage{graphicx}
\usepackage{bm}
\usepackage{amsmath} 

\begin{document}



\title{Solitons in a discrete model of chiral liquid crystals with competing interactions}

\author{Alain M.~Dikand\'e} \email{dikande.alain@ubuea.cm}
\affiliation{Laboratory of Research on Advanced Materials and Nonlinear Science (LaRAMaNS), Department of Physics, Faculty of Science, University of Buea P. O. Box 63 Buea, Cameroon}  

\author{Bernard Y. Nyanga} 
\affiliation{Department of Physics, Faculty of Science, University of Buea PO Box 63 Buea, Cameroon}

\author{S. E. Mkam Tchouobiap}
\affiliation{ Laboratory of Research on Advanced Materials and Nonlinear Science (LaRAMaNS), Department of Physics, Faculty of Science, University of Buea P. O. Box 63 Buea, Cameroon}
%
%

\date{\today}

\begin{abstract}
Chiral liquid crystals exhibit in-plane spontaneous polarizations, however in their smectic phase the primary order parameter is a tilt vector associated with molecular rotations around the long molecular axis parallel to the director. The molecular rotations lead to several distinct phases among which a domain-wall texture with a periodic-kink soliton profile. In this study the formation of domain walls in smectic chiral liquid crystals is analyzed, with emphasis on the competition between ising-type symmetric and antisymmetric nearest-neighbor interactions, and an in-plane electric field. It is found that antisymmetric intermolecular interactions, which are of chiral origin, increase the width of kink structures in the domain wall at moderate intensity of the $y$ component of the electric field. Increasing the $x$ component of the electric field creates unstable condition for soliton formation irrespective of magnitudes of the symmetric and chiral intermolecular interactions. Stability condition for single-kink domain-wall structures in the discrete molecular chain, is discussed by estimating the Peierls stress experienced by the single-kink soliton.  Results suggest that chirality lowers the Peierls-Nabarro barrier, hence increasing the lifetime of single-kink structures in the discrete medium. 
\end{abstract}

\pacs{42.60.Da, 42.65.Sf, 42.65.Tg, 05.45.Pq}
\maketitle


\section{\label{sec:1} Introduction}
Discrete linear (i.e. one-dimensional) chains exhibit a broad range of phase transitions resulting from the competition between different interactions among atoms or molecules \cite{1}. The soliton condensation, one of most intriguing of these phases, occurs as a disclination due to violation of rotational symmetries in molecular crystals \cite{1,2,3,4,5}, as dislocation in crystal lattices due to misalignement of atoms in the crystal structure of a Frenkel-Kontorova lattice \cite{1,7,8,9}, as a discommensuration associated with a superlattice structure forming in a charge-density-wave, antiferromagnetic or ferromagnetic lattices \cite{10a,10,11,12,13,14}, or as domain walls in incommensurate systems in general. In these systems solitons are topological defects formed from the presence of different nonlinear interactions competing with the chain discreteness and, in some physical contexts, an external field. In mathematical physics they are solutions to nonlinear partial differential equations, for which they represent waves of long lifetime consequent upon the balance of lattice dispersion by nonlinearity. \\
Among physical systems exhibiting the soliton condensation phase are a class of soft matters composed of non-spherical molecules, known to be prone to structural orders governed by molecular tilts with respect to the molecular-chain axis \cite{15,16}. In this class liquid crystals \cite{lam} have attracted a great deal of attention because of the elongated (i.e. rod-like) shape of their molecular constituants \cite{17,18,18b,19}, this strong anisotropy of molecules indeed favors phase transitions related to the molecular tilts leading to several distinct phases in liquid crystals. \\
In the smectic (Sm) phase, liquid crystals possess non-zero molecular tilt angles between the director and the normal to the smectic layers. In general there exists two distinct tilt directions i.e. a parallel and an antiparallel tilt directions, associated with two possible orders i.e. the ferroelectric order, in which molecules are tilted in unison with an average zero phase difference between tilts of neighbor molecules, and the antiferroelectric order in which phase differences fluctuate around $\pi$ \cite{18,18b,18a,19a,19b,20,21,22,23,24,25}. When molecules have no internal planes of symmetry, the pitch becomes chiral and molecular tilts cause helicoidal distorsions in which molecular rotations around the layer axis are strongly biased. This leads to a frustration in interayer interactions, and in-plane polarizations precessing around the helical axis from a layer to neighbour layers. In the presence of an external field the precessions of molecular polarizations are confined within the smectic layers thus triggering a long-range orientation order via a Freedericksz transition \cite{fred,fred1}. \\
In real Sm chiral (SmC) liquid crystals, however, the directions of in-plan polarizations on neighboring smectic layers are neither exactly parallel nor exactly antiparallel. Actually the helical superstructure \cite{18} is stabilized by a competition between the frustration caused by chirality and the ising-type coupling promoting ferroelectric or antiferroelectric orders along the tilt axis. Two important consequences of such competition are a slow precession of in-plane electrical polarizations \cite{18,18a,21,22,23,24,25}, and the possibility of an electroclinic effect \cite{greca,grec,nysh} such as chiral piezoelectricity, a perpendicular alignement of the tilt direction with in-plane polarizations or with an applied external field \cite{greca,grec,nysh,musa,musb}. \\
In this work, we consider a discrete model for chiral SmC liquid crystals in which both the symmetric and antisymmetric intermolecular interactions are taken into account. We also take into consideration electroclinic effects via a cross coupling between the tilt order parameter and an applied electric field, assumed to describe a perpendicular alignement of the tilt direction with an applied planar electric field \cite{grec}. From the proposed discrete model we investigate the effects of the competition between the symmetric (i.e. ising-type) and antisymmetric (i.e. chiral-type) intermolecular interactions, and the electric field confined within the smectic layers, on the generation and shape profiles of domain-wall structures in the chiral SmC liquid crystal. Under specific conditions the model reduces to the standard sine-Gordon equation without suppression of chiral intermolecular interactions, and hence can support sine-Gordon kink solitons. Since the discrete equation is not integrable, we resort to a continuum approximation wich requires an analysis of the effect of lattice discreteness on continuum soliton profiles. In this respect we carry out numerical silumations which show that the continuun periodic-kink soliton has the same profile as the exact numerical solution to the discrete problem, but not the continuum single-kink solution. Therefore the discreteness effect will be more effective on the single-kink soliton, and we determine the Peierls stress experienced by this structure in the discrete model. We discuss the implictions of the variations of the Peierls-Nabarro barrier with the chiral and ising-type intermolecular interaction coefficients, and magntitudes of the two components of the external field, on the single-kink lifetime in the discrete molecuar medium.

\section{\label{sec:two} The discrete model and soliton structures}
\begin{figure*}
\centering
\begin{minipage}{0.5\textwidth}
\includegraphics[width=3.in, height=2.65in]{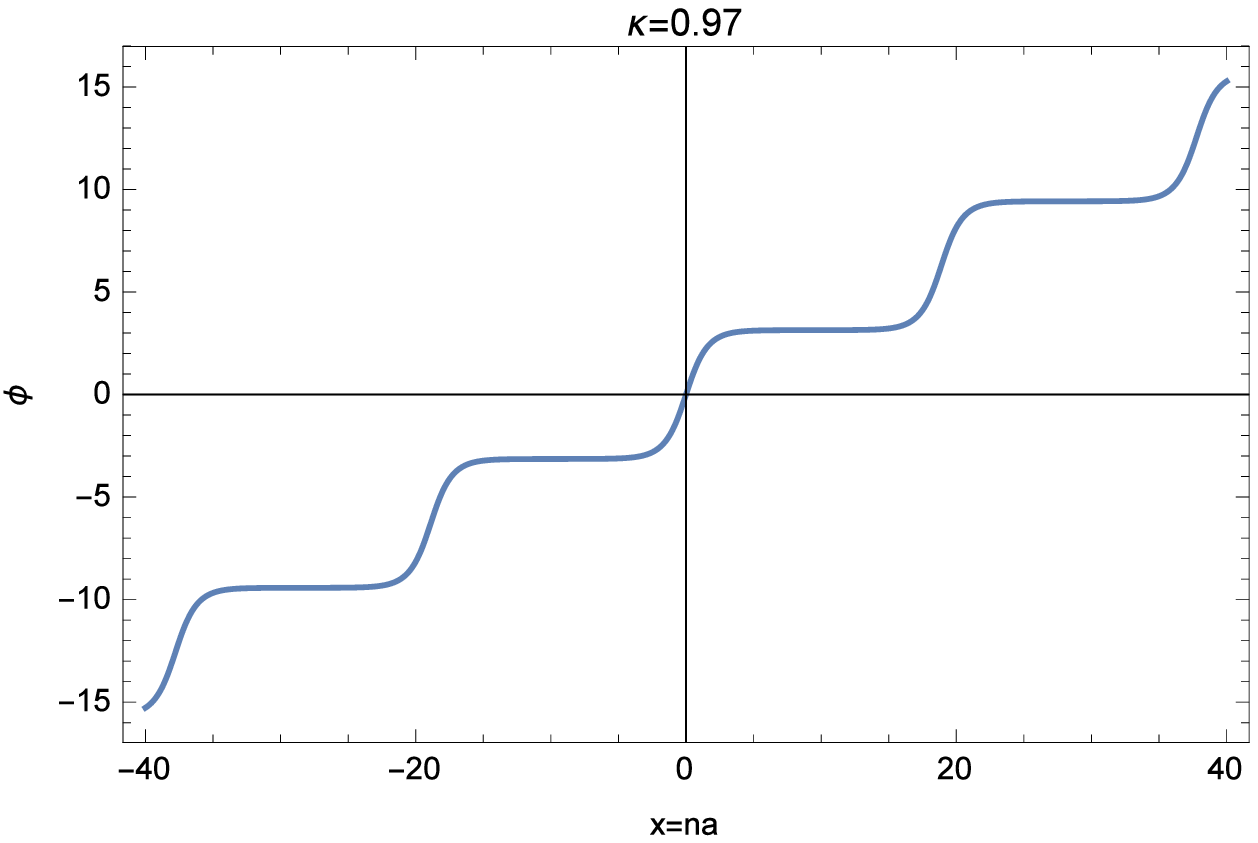}
\end{minipage}\hfill
\begin{minipage}{0.5\textwidth}
\includegraphics[width=3.in, height=2.65in]{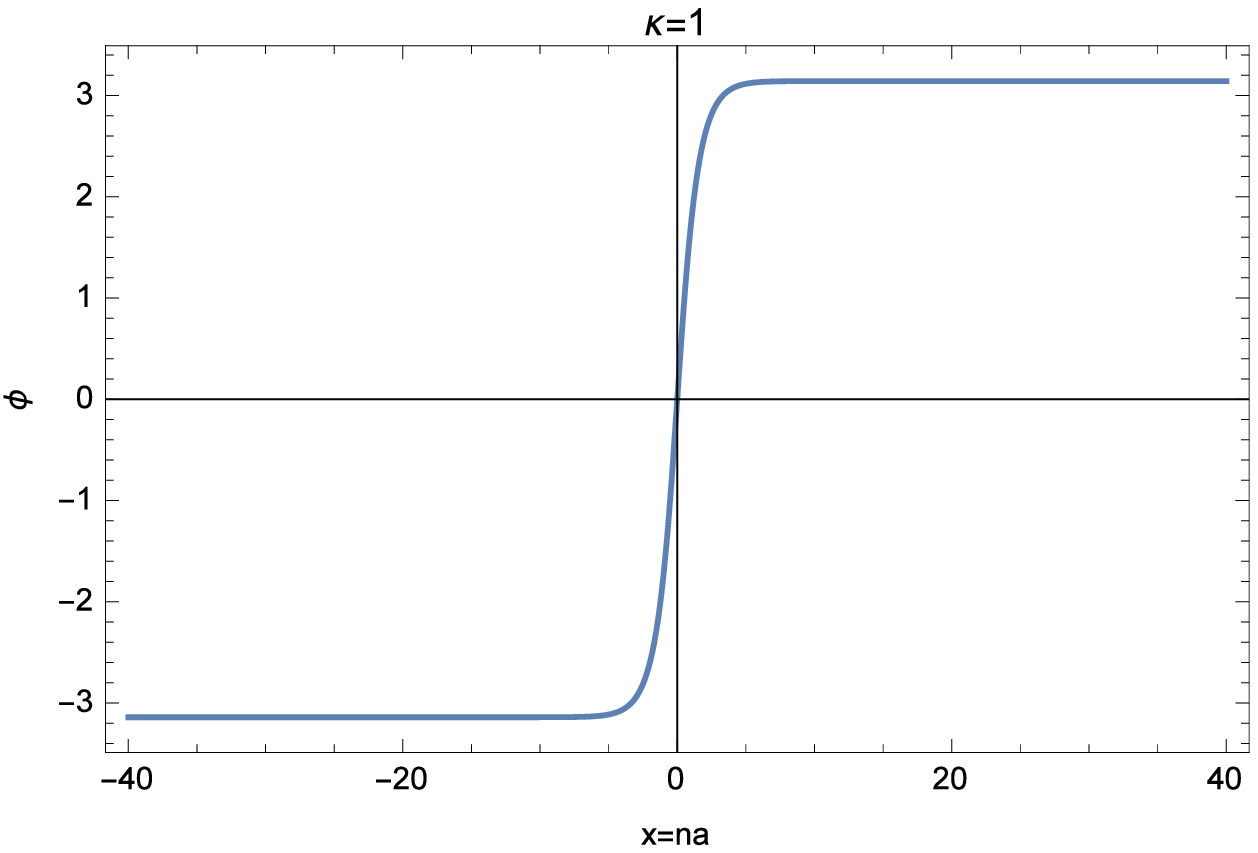}
\end{minipage}
\caption{\label{fig1} (Color online) Saptial profile of the analytical periodic-kink soliton solution (\ref{eq9}), for $\kappa=0.97$ (left graph) and $\kappa=1$ (right graph). All other parameters are taken to be unity.}
\end{figure*}

\begin{figure}[h]
\includegraphics[scale=.5]{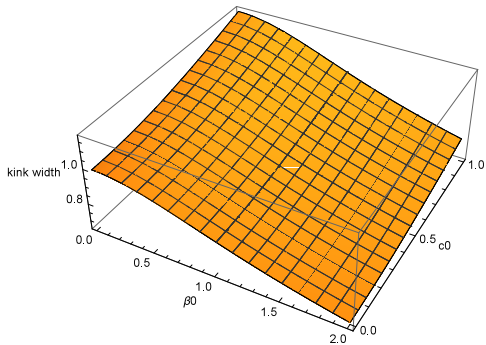}
\caption{\label{fig2}
(Color online) Width of kinks (in units of the lattice spacing) in the periodic-kink solution, plotted versus $\beta_0$ and $c_0$ for a fixed value of $\varepsilon_y$ (i.e. $\beta_0=1$).} 
\end{figure}

Our model is a discrete linear chain of rod-shaped molecules describing a liquid crystal in the SmC phase. In this phase the tilt of the director from the normal (here the z axis) to the smectic layers (xy plane) in the $n^th$ smectic layer is a two-component vector field $\mathbf{u}_n$, representing the magnitude and direction of the tilt. We consider the system in the vicinity of the smectic-A to the tilted (i.e. SmC) phase transition, and assume that in addition to achiral nearest-neighbor interactions molecules are also antisymmetrically coupled, as a result of chiral interactions between molecules on neihgbor smectic layers. Due to the chirality the tilt direction will tend to align perpendicular to any applied electric field within the smectic layer by virtue of the electroclinic effect. Taking this last effect into consideration, the discrete Landau-Ginzburg free energy corresponding to our model will be \cite{sep,skar}:
\begin{eqnarray}
G&=&\sum_{n=1}^N\lbrack\frac{A}{2}\mathbf{u}_n^2 + \frac{B}{2}\mathbf{u}_n^4 + J\mathbf{u}_n. \mathbf{u}_{n+1} \nonumber \\
&+& f(\mathbf{u}_n \times \mathbf{u}_{n+1})_z + (\mathbf{E}\times\mathbf{u}_n)_z \rbrack. \label{eq1}
\end{eqnarray}
In Eq. (\ref{eq1}) the parameter $A(T)=A_0(T-T_c)$, where $A_0$ is positive ensuring a continuous transition from the smectic A phase to the SmC phase at the mean-field critical temperature $T_C$. $B$ in the second term is positive, the third term is an ising-type symmetric nearest-neighbor interactions while the four term is an antisymmetric nearest-neighbor interaction due to molecular chirality. The last term takes into account the electroclinic effect caused by chirality, which forces the directions of the tilt vector and an applied electric field $\mathbf{E}$ to be perpendicular \cite{grec}. The supscript $z$ in Eq. (\ref{eq1}) indicates a projection along the smectic layer normal $z$. \\
Express the local order parameters $\mathbf{u}_n$ as two-component verctor fileds i.e. $\mathbf{u}_n\equiv u_0\,(\cos\varphi_n, \sin \varphi_n)$, where the tilt amplitude $u_0$ is assumed homogeneous and only the angle of helicoidal motion $\varphi_n$ varies locally. As for the applied electric field, we assume that its lies within the layer planes and hence is a two-component vector $\mathbf{E}\equiv (E_x, E_y)$. In terms of $\varphi_n$ Eq. (\ref{eq1}) becomes:  
\begin{eqnarray}
G&=&G_0 + \sum_{n=1}^N\lbrack J u_0^2\cos(\varphi_{n+1} -\varphi_n) + f u_0^2\sin(\varphi_{n+1} -\varphi_n)\nonumber \\ 
&-&(E_x\sin \varphi_n - E_y\cos\varphi_n)\rbrack. \label{eq2}
\end{eqnarray}
The spatial configuration of the helicoidal order in the chiral SmC phase is obtained by minimizing (\ref{eq2}) with respect to $\varphi_n$, and is governed by the discrete equation:
\begin{eqnarray}
0&=&\sin(\varphi_{n+1} -\varphi_n) + \sin(\varphi_{n-1} -\varphi_n)- c_0 \lbrack\cos(\varphi_{n+1} -\varphi_n) \nonumber \\
&+& \cos(\varphi_{n-1} -\varphi_n)\rbrack +\epsilon_x\cos \varphi_n + \epsilon_y\sin\varphi_n, \label{eq3} 
\end{eqnarray}
where:
\begin{equation}
c_0= \frac{f}{J}, \hskip 0.3truecm \epsilon_x=\frac{E_x}{J u_0^2}, \hskip 0.3truecm \epsilon_y=\frac{E_y}{J u_0^2}. \label{eq4}
\end{equation}
To solve Eq. (\ref{eq3}) we will isolate $\varphi_{n+1}$ from the local variables $\varphi_n$ and $\varphi_{n-1}$. To this end we define:
\begin{equation}
\beta_0= \frac{\epsilon_x}{\epsilon_y}, \hskip 0.3truecm \alpha_0= \epsilon_y\sqrt{\frac{1+\beta_0^2}{1+c_0^2}}, \label{eq5}
\end{equation}
such that Eq. (\ref{eq3}) reduces to:
\begin{widetext}
\begin{equation}
\varphi_{n+1} = \varphi_n + \arctan c_0 + \arcsin\left[ \sin(\varphi_n -\varphi_{n-1} -\arctan c_0) - \alpha_0\sin(\varphi_n + \arctan \beta_0)\right]. \label{eq6} 
\end{equation}
\end{widetext}
The equilibrium solutions of Eq. (\ref{eq6}) will generally depend on the signs and magnitudes of characteristic parameters of the model. The equilibrium states for instance, in the absence of applied electric field (i.e. $\mathbf{E}=0$), have been discussed in some past works considering second-neighbor interactions of both achiral and chiral types \cite{sep,skar}. Thus, it is well established that an antiferroelectric groundstate is expected mainly when second-nearest neighbor interactions are taken into account\cite{musb}. From the standpoint of Eq. (\ref{eq6}) without the external field but with addition of second-nearest neighbor interactions, this state will correspond to an equilibrium configuration where the phase differences between tilt directions increase nearly by $\pi$ \cite{musb}. In the present context, where there is no second-nearest neihghbor interaction terms in Eq. (\ref{eq6}), we can rule out antiferroelectric ordering. This is anyhow evident given that the only equilibruim state suggested by Eq. (\ref{eq6}), when $\mathbf{E}=0$, is the zero phase difference between neighbor tilts along the chain axis corresponding to a ferroelectric order. \\ Looking for the general solution of Eq. (\ref{eq6}) for arbitrary values of the model parameters, it is instructive recalling that a similar equation was obtained in the study of discrete-soliton and soliton-lattice generations during the unwinding process of $SmC^*_{\alpha}$ phase to SmC phase driven by an electric field \cite{yuk}. Eq. (\ref{eq6}) is more pricisely a perturbed version of the so-called sine-lattice equation \cite{9,tak1,tak2}, and as such is not exactly integrable. However, in some specific contexts approximate solutions can be found. For instance, when $c_0=\beta_0=0$, a family of approximate solutions have been shown to exist with some dispersion relation \cite{9,tak1,tak2}. These solutions are $\pi$-kink solitons and are also solutions to the discrete equation (\ref{eq6}) with $c_0=0$. Indeed, with the variable change $\varphi_n=\phi_n-\arctan \beta_0$ Eq. (\ref{eq6}) can be transformed to \cite{9}:
\begin{equation}
\phi_{n+1} = \phi_n + \arcsin\left[\sin(\phi_n -\phi_{n-1}) - \alpha_0\sin\phi_n\right]. \label{eq7} 
\end{equation}
The single-soliton solution to the sine-lattice equation (\ref{eq7}) is a $\pi$ kink as shown in ref. \cite{tak2}, with the help of Hirota transformations. \\
The single-kink soliton as a general solution to the sine-lattice equation is interesting, but concerning the specific problem at hand periodic structures provide a better picture of the topology of the helicoidal superstructure created in the discrete system. In want of analytical method enabling the derivation of an exact solution consistent with this periodic helicoidal superstructure, we shall resort to a continuum-limit approximation. In this goal we assume the phase differences $\phi_{n+1}-\phi_n$ to remain always small, such that we can expand $\sin(\phi_{n+1}-\phi_n)\approx \phi_{n+1}-\phi_n$, $\sin(\phi_n-\phi_{n-1})\approx \phi_n-\phi_{n-1}$. Substituting these expansions in Eq. (\ref{eq6}) and defining a continuous spatial position $x=na$, where $a$ (hereafter assumed to be unity) is the separation between neighbor smectic layers at equilibrium, we can readily rewrite (\ref{eq6}) as:
\begin{equation}
\phi_{xx} = -\alpha_0\,\sin \phi, \label{eq8} 
\end{equation}
where the subscript "$xx$" means a second-order derivative with respect to $x$. The periodic-soliton solution to Eq. (\ref{eq8}) is obtained as: 
\begin{eqnarray}
\phi_{\kappa}(x) &=& \pm 2\arcsin \left[ sn\left(\frac{x}{\ell_{\kappa}}, \kappa \right)\right], \nonumber \\
&=& \pm 2\,am\left(\frac{x}{\ell_{\kappa}}, \kappa \right), \hskip 0.2truecm \ell_{\kappa}= \frac{\kappa}{\sqrt{\alpha_0}}, \label{eq9} 
\end{eqnarray}
in which $sn$ and $am$ are Jacobi elliptic functions of modulus $\kappa$ obeying $0\leq \kappa \leq 1$ \cite{abra,dika,dikb,dikc}. Explicitely the solution (\ref{eq9}) describes a lattice of identical kinks of equal width $\ell_{\kappa}$ and equal separation $d_{\kappa}= 2\ell_{\kappa}K(\kappa)$, where $K(\kappa)$ is the complete elliptic integral of the first kind. \\
According to the expression of $\ell_{\kappa}$ given in formula (\ref{eq9}), an increase of the kink width with increase of the chiral interaction strength at fixed value of the symmetric interaction $J$, is balanced by an increase of the $x$ component of the electric field for a fixed value of $\epsilon_y$. However, this balance costs a uniform shift of the periodic-kink soliton by a phase factor $\arctan(\beta_0)$, as reflected in the expression of the real solution to our problem i.e. $\varphi(x)= \phi(x) - \arctan \beta_0$. Variations of the period $d_{\kappa}$ with $c_0$ and $\beta_0$, are the same as the variations of the kink width with these two parameters. In fig. \ref{fig1}, we plotted the amplitude-function solution given by (\ref{eq9}) for $\kappa=0.97$ (left graph) and $\kappa=1$ (right graph). 
It is remarkable that profile of the amplitude-function solution when $\kappa=1$, coincides with the analytical expression:
\begin{equation}
\phi(x)=\pm 4\arctan\left[ \exp{\frac{x}{\ell}}\right] \mp \pi, \label{1sol}
\end{equation}
while the period $d_{\kappa=1}\rightarrow \infty$. Clearly, when $\kappa\rightarrow 1$, the helicoidal superstructure decays to a single-kink soliton. Fig. \ref{fig2} summarizes the variations of the periodic-kink width with $\beta_0$ and $c_0$, for $\kappa=0.97$.
\par
\begin{figure*}
\centering
\begin{minipage}{0.5\textwidth}
\includegraphics[width=3.in, height=2.65in]{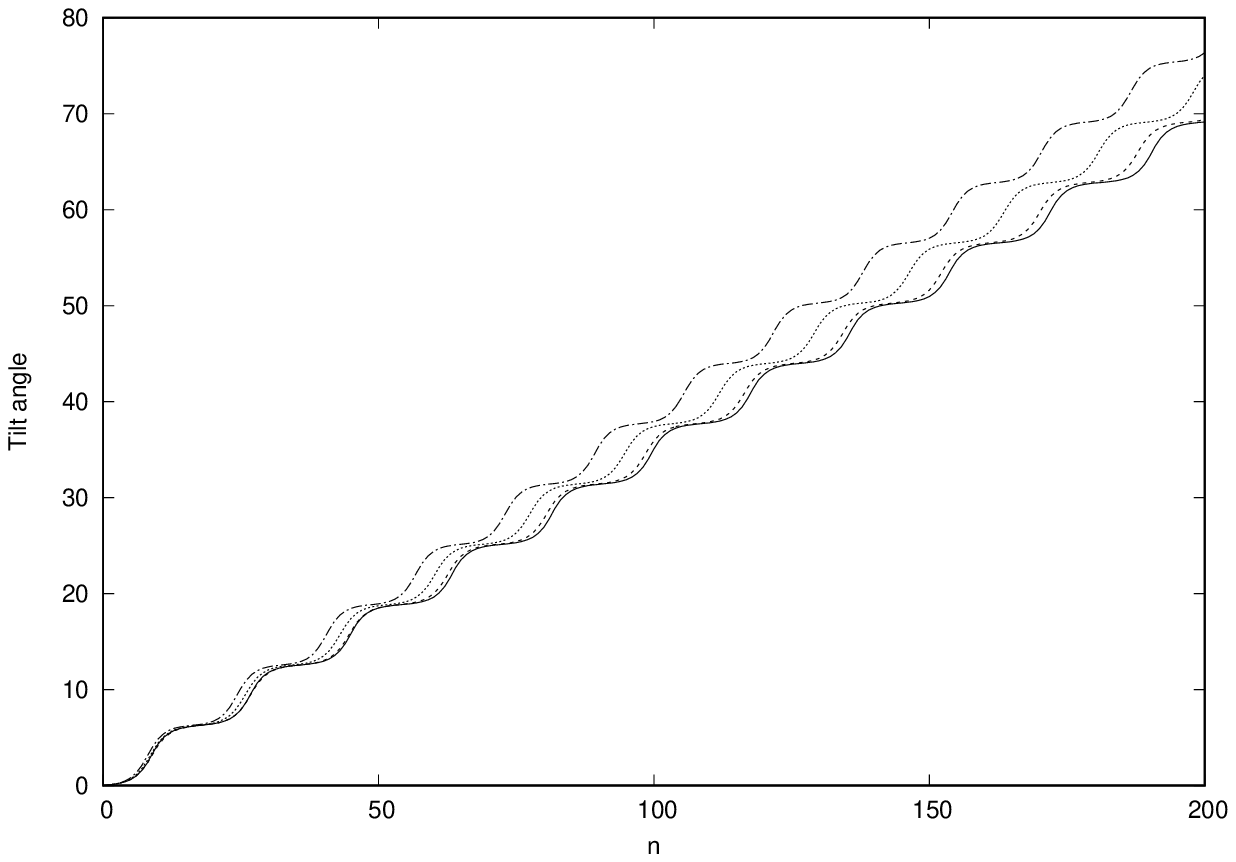}
\end{minipage}\hfill
\begin{minipage}{0.5\textwidth}
\includegraphics[width=3.in, height=2.65in]{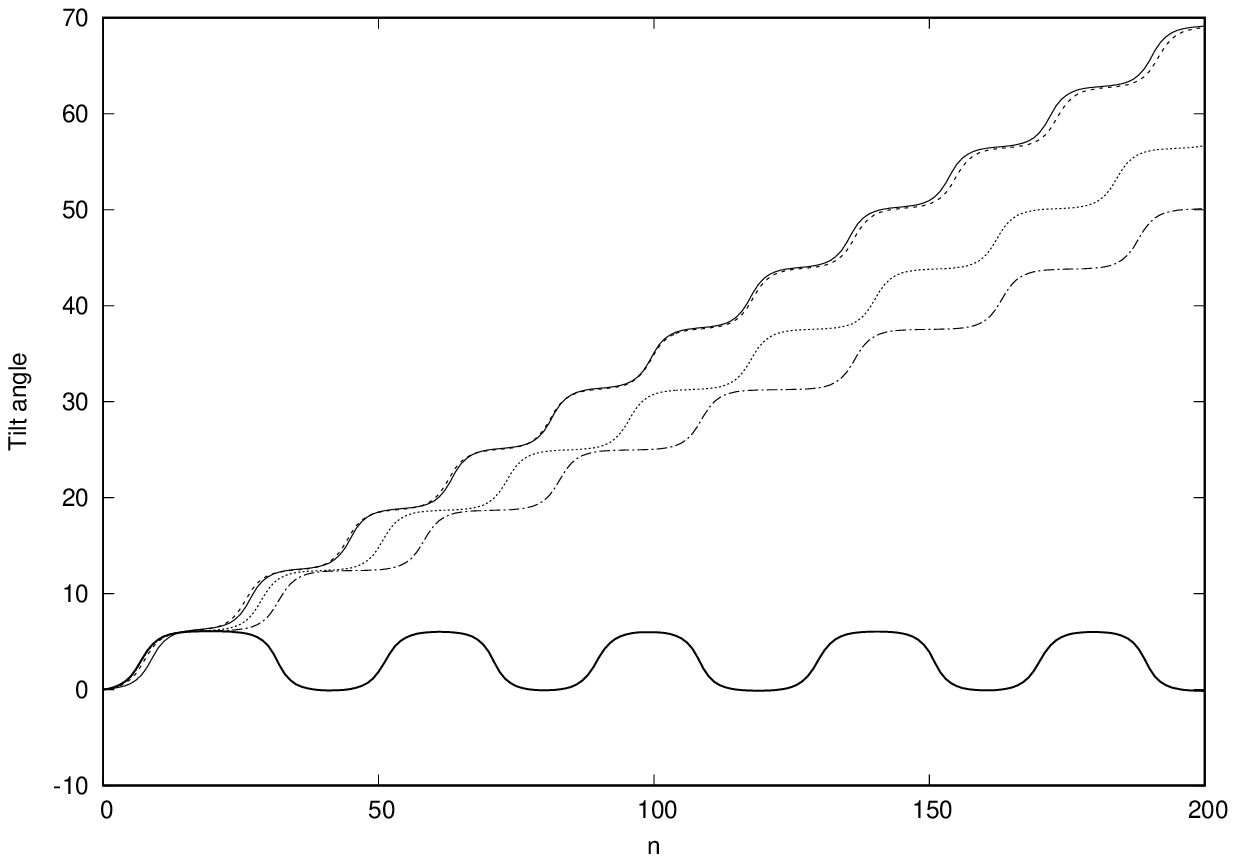}
\end{minipage}
\caption{\label{fig3} (Color online) Spatial profiles of the periodic-kink soliton from numerical simulations ofthe discrete equation  (\ref{eq6}), for $\varepsilon=0.2$. Left panel: $\beta_0=0$ (fixed) and $c_0=0$, $0.2$, $0.5$ and $0.75$ (from bottom to top curves). Right panel: $c_0=0.1$ (fixed), $\beta_0=0$, $0.1$, $0.16$, $0.17$ and $0.176$ (from top to bottom curves).}
\end{figure*}

The above results suggest the existence of two possible distinct kink solitons for the continuum problem, namely the single-kink and periodic-kink solitons. To identify which of these two solutions is closer to the exact solution to the discrete problem, and hence to check the consistent of the continuum limit approximation, we simulated numerically the full disrete equation (\ref{eq6}) (using a Fibonacci-type algorithm). In the simulations we fixed $\varepsilon_y$ to $0.2$ (an arbitrary value), and varied $c_0$ and $\beta_0$. Curves in the left graph of Fig. \ref{fig3} are spatial profiles of the periodic-kink soliton when $\beta_0=0$ and $c_0=0$, $0.2$, $0.5$, $0.75$. In the right panel, the periodic-kink soliton profiles were generated for a fixed value of $c_0$ (i.e. $c_0=0.1$) while $\beta_0$ was varied as $\beta_0=0$, $0.1$, $0.16$, $0.17$ and $0.176$.
As the left graph clearly suggests, periodic-kink tructures are shape preserving and always stable in the discrete system even for an electric field reduced to only its $y$ component. As we increase the chiral interaction coefficient (which corresponds to increasing $c_0$) kinks in the soliton lattice get sharper while their widths are increased, consistent with the behavior observed in Fig. \ref{fig2}. On the other hand, when the ratio of the chiral to the symmetric intermolecular interactions is fixed, and the $x$ component of the electric field is increased from a zero, the periodic-kink profile seems to survive only up to some critical value of the ratio $E_x/E_y$ as seen in the right graph of Fig. \ref{fig3}. Beyond this critical value the soliton structure decays into a kink-antikink lattice, which in turn will survive within some finite range of values of $\beta_0$ beyond which soliton structures become unstable in the system.           
\section{\label{sec:three} Energetic considerations on soliton existence and stability}
Energies of solitons are relevant parameters when considering their formation as well as their stability in a given medium. For the problem at hand there are two different energies that are relevant for the existence of the kink solitons obtained in the previoeus section, they are their creation energy and the energy related to the discreteness of the molecular chain. Numerical simulations of the discrete equation ({\ref{eq6}) carried out in sec. \ref{sec:two}, have shown that profiles of the exact periodic-kink solutions to this equation were identical with the continuum periodic-kink soliton solution obtained in formula (\ref{eq9}). This means Eq. (\ref{eq9}) can be readily regarded as an exact solution to the discrete equation, and too is its continuum energy. On the contrary, the single-kink solution Eq. (\ref{1sol}) is not reproduced by numerical simulations and hence can by no means be exact to the discrete problem. Reason why the single-kink soliton Eq. (\ref{1sol}) is expected to suffer the lattice discreteness, resulting in energy dispersion in efforts to overcome the discrete relief of the molecular chain. To start we calculate the creation energy of the periodic-kink soliton, and next determine the potential barrier erected by the lattice discreteness and to which the single-kink soliton can be trapped. \\
The periodic-kink soliton solution (\ref{eq9}) was obtained by integration of Eq. (\ref{eq8}) with periodic boundary conditions. The corresponding energy integral, in the specific case of Jacobi-elliptic function solutions, is given by: 
\begin{eqnarray}
\frac{1}{2}\phi_x^2&=& \frac{\omega_0^2}{\kappa^2}\left[1-\kappa^2\sin^2(\phi/2)\right], \nonumber \\
 &=&\omega_0^2V(\phi), \hskip 0.2true cm \omega_0^2=\alpha_0. \label{fi}
\end{eqnarray}
This relation provides the right condition of energy conservation for the periodic-kink soliton, and hence can be used to define the express of the total energy of the periodic-kink soliton i.e.: 
\begin{equation}
E_{sol}(\kappa)= J u_0^2\,\int_0^{d_{\kappa}}{dx\left[\frac{1}{2}\phi_x^2 + \omega_0^2V(\phi)\right]}. \label{fi1}
\end{equation}
Substituting $\phi(x)$ given by (\ref{eq9}) in the last integral we find:
\begin{equation}
E_{sol}(\kappa)= J u_0^2\frac{4\sqrt{\epsilon_y}}{\kappa}\left(\frac{1+\beta_0^2}{1+c_0^2}\right)^{1/4}E(\kappa), \label{ensol}
\end{equation}
where $E(\kappa)$ is the complete elliptic integral of the second kind \cite{abra}. Formula (\ref{ensol}) indicates that an enhancement of chirality will increase the periodic-kink creation energy, while an increase of the $x$ component of the electric field will be detrimental to the creation of periodic-kink soliton. This behavior, once again, is consistent with results of numerical simulations of the discret equation (\ref{eq6}) discussed in the previous section. In the single-kink limit formula (\ref{ensol}) reduces to:
\begin{equation}
E_{sol}= 4J u_0^2\omega_0. \label{ensol1}
\end{equation}
As emerged in our previous discussions, strickly formula (\ref{ensol1}) is valid only in the continuum medium given that the single-kink structure is not exact for the discrete system, to find the actual energy of the single-kink solution placed in the discrete molecular chain, we must use the analytical solution (\ref{1sol}) in the discrete total energy given by (\ref{eq2}). To this aim we must explicely introduce a pinning coordinate for the single-kink soliton, here denoted $X$ and coinciding with the soliton centre-of-mass position \cite{wil} in the discrete discrete system. Thus the argument of (\ref{1sol}) is shifted from $x=n$ to $x-X$. Next keeping the "ferroelectric-ordering" argument i.e. $\phi_{n+1}-phi_n$ is always very small, using formula (\ref{1sol}) and grouping all constant terms in an homogeneous function $F_0$, the discrete energy (\ref{eq2}) can be written:  
\begin{eqnarray}
F&=&F_0 - 6Ju_0^2\sqrt{1+c_0^2}\,\alpha_0\,\sum_{n=1}^N{sech^2\left(\frac{n-X}{\ell}\right)} \label{eq1a} \\
&+& 4Ju_0^2\sqrt{\alpha_0(1+c_0^2)}\,\arctan c_0 \sum_{n=1}^{N}{sech\left(\frac{n-X}{\ell}\right)}, \nonumber
\end{eqnarray}
where $F=G_0-G$. The discrete sum over $n$ in (\ref{eq1a}) is exact when $N\rightarrow \infty$, yielding:
\begin{eqnarray}
F&=&U_0 + 4Ju_0^2\sqrt{1+c_0^2}[dn(2X K(\nu))\arctan c_0 \nonumber \\
&-& 3\left(K(\nu)dn^2(2X K(\nu)) - E(\nu)\right)] K(\nu), \nonumber \\
U_0&=&F_0 - 24Ju_0^2\sqrt{\alpha_0(1+c_0^2)}. \label{enar}
\end{eqnarray}
$dn()$ is one of Jacobi elliptic functions \cite{abra} while $K(\nu)$ and $E(\nu)$ are complete elliptic integrals of the first and second kinds respectively, here given in terms of a new modulus $\nu$ obeying the transcendental relation \cite{mal}: 
\begin{equation}
\pi \ell=\frac{K(\nu')}{K(\nu)}, \hskip 0.25truecm \nu'=\sqrt{1-\nu^2}, \hskip 0.25truecm 0\leq \nu\leq 1. \label{enar1}
\end{equation}
To easily capture the physics in the expression (\ref{enar}) of the discete energy, we adopt the Fourier series representations of the Jacobi elliptic functions $dn()$ and $dn^2()$ \cite{abra} and find: 
\begin{eqnarray}
F&=& U_0 + \sum_{p=1}^{\infty}{U_p(\ell)\,\cos(2\pi pX)}, \label{four} \\
U_p(\ell)&=&4\pi Ju_0^2\sqrt{1+c_0^2}\left[\frac{6p\pi}{\sinh\left(p\pi^2\ell\right)} - \frac{\arctan c_0}{\cosh\left(p\pi^2\ell\right)}\right], \nonumber 
\end{eqnarray}
where the transcendental relation (\ref{enar1}) was used to eliminate the complete elliptic integrals $K(\nu)$ and $E(\nu)$. According to formula (\ref{four}), in the discrete regime the single-kink soliton energy is a periodic function of the soliton centre-of-mass position $X$ with an energy amplitude $U_p(\ell$. When the kink width $\ell$ is large enough, the leading term $U_1(\ell)$ in the sum (\ref{four}) will dominate and the amplitude of the periodic energy reduces to:
\begin{equation}
U_{PN}=4\pi Ju_0^2\sqrt{1+c_0^2}\left[\frac{6\pi}{\sinh\left(\pi^2\ell\right)} - \frac{\arctan c_0}{\cosh\left(\pi^2\ell\right)}\right]. \label{vpn}
\end{equation}
$U_{PN}$, which we refer to as the Peierls-Nabarro barrier, is the amplitude of the periodic potential experienced by the single-kink soliton due to the discreteness of the molecular chain. Instructively formula (\ref{vpn}) reveals that the Peierls-Nabarro barrier will be lowered by the chirality, while the contribution from the electric field is a decrease of the kink width and hence an increase of the Peierls-Nabarro barrier. However the dependence of $U_1$ in both the chirality and the electric field, reflected by formula (\ref{vpn}, is such that the chirality will have the dominant effect for a decrease of $\ell$ with an increase of $\beta_0$, will be balanced by the increase of $c_0$.   
\section{\label{sec:four} Conclusion}
We investigated the effects of the competition between an ising-type nearest-neighbor interaction and an antisymmetric nearest-neighbor interaction (of chiral origin) between molecules on one hand, and a two-component electric field on the other hand, on the formation and stability of domain walls in chiral smectic liquid crystals in the ferroelectric phase. We found that the equilibrium configuration of the discrete liquid-crystal system, resulting from molecular tilts with respect to the long molecular chain axis, is described by a sine-lattice type equation. In the continuum limit this equation can be reduced to the classic sine-Gordon equation, thus admitting two distinct soliton solutions namely a single-kink and kink-lattice (i.e. periodic-kink) soliton solutions. While numerical simulations of the full sine-lattice equation suggest that the continuum periodic-kink solution can be a good approximation of the exact solution to the discrete problem, the single-kink solution can by no mean be obtained from the discrete equation and therefore remains exact only in the continuum limit. Nevertheless, given that a long-range domain-wall order forms by nucleations of single-kink soliton structures, we considered the survival of such structures in the discrete molecular chain. In this respect we obtained the amplitude of the Peierls-Nabarro potential, which inverse is proportional to kink lifetime in the presence of lattice discreteness, and obtained that chirality lowers the Peierls stress and consequently favors the single-kink stability in the discrete system.\\
In this study we were concerned mainly with the competing effects of the ising-type and antisymmetric intermolecular interactions, as well as the electric field, on the formation of solitonic structures in chiral smectic liquid crystals. Although the in-plane polarizations are secondary order parameters, and hence were not considered in this work, it is well established \cite{musb} that because of the chirality in-plane polarizations of molecules are not parallel with the primary order paramater (i.e. the tilt vector ${\bf u}$). Therefore the interaction of in-plane polarization vectors and the tilt vectors will introduce Lifshitz terms in the total energy accounting for a chiral piezoelectric effect \cite{musb,sep}). A study of the formation of domain walls taking into account this chiral piezoelectric effect is a relevant open problem, which will certainly provide rich insight onto the physics of discrete smectic chiral liquid crystals with competing interactions.   

\acknowledgments
A. M. Dikand\'e whishes to thank the Alexander von Humboldt (AvH) foundation for logistic supports. The laboratory of Research on Advanced Materials and Nonlinear Sciences (LaRAMaNS) is partially funded by the World Academiy of Sciences (TWAS), Trieste, Italy.

\end{document}